\documentclass[12pt,oneside]{article}
\usepackage{epsfig, cite}
\usepackage{color}
\usepackage{ifthen}
\usepackage{graphicx}

\def\({\left(}
\def\){\right)}
\def\[{\left[}
\def\]{\right]}

\def\e{\begin{equation}}
\def\q{\end{equation}}
\def\m{\begin{eqnarray}}
\def\n{\end{eqnarray}}

\begin{document}
\thispagestyle{empty} \setcounter{page}{0}

\vspace{2cm}

\begin{center}
{\huge Weak Gravity Conjecture with Large Extra Dimensions}

\vspace{1.4cm}

Qing-Guo Huang

\vspace{.2cm}

{\em School of physics, Korea Institute for
Advanced Study,} \\
{\em 207-43, Cheongryangri-Dong,
Dongdaemun-Gu, } \\
{\em Seoul 130-722, Korea}\\
\end{center}

\vspace{-.1cm}

\centerline{{\tt huangqg@kias.re.kr}} \vspace{1cm}
\centerline{ABSTRACT}
\begin{quote}
\vspace{.5cm}

In the presence of large extra dimensions, the fundamental Planck
scale can be much lower than the apparent four-dimensional Planck
scale. In this setup, the weak gravity conjecture implies a much
more stringent constraint on the UV cutoff for the U(1) gauge theory
in four dimensions. This new energy scale may be relevant to LHC.

\end{quote}
\baselineskip18pt

\noindent

\vspace{5mm}

\newpage

\setcounter{equation}{0}

Extra dimensions are naturally required in some fundamental theory.
For example, the self-consistency condition for superstring theory
requires that the critical dimension should be ten
\cite{Polchinski:1998rq}; otherwise, the conformal anomaly on the
world sheet can not be canceled exactly. A new higher-dimensional
mechanism for solving the hierarchy problem was proposed in
\cite{Randall:1999ee}. Compactification from higher dimensions to
four dimensions not only is necessary for connecting string theory
with our real world, but also gives us many surprising insights.

In \cite{Vafa:2005ui}, Vafa pointed out that gravity and the other
gauge forces can not be treated independently and the vast series of
semi-classically consistent effective field theories are actually
inconsistent after gravity is included. Furthermore, the authors of
\cite{Arkani-Hamed:2006dz} proposed the weak gravity conjecture
which can be most simply stated as gravity is the weakest force.
This conjecture implies that in a four-dimensional theory with
gravity and a U(1) gauge theory, there is a new intrinsic UV cutoff
\e \Lambda\sim gM_4 \label{lgm}\q induced by four-dimensional
gravity characterized by the four-dimensional Planck scale $M_4$,
where $g$ is the gauge coupling. Some related topics are discussed
in
\cite{Huang:2006hc,Huang:2006tz,Li:2006jj,Adams:2006sv,Kachru:2006em,Li:2006vc,Ooguri:2006in,Kats:2006xp,Banks:2006mm,Gogoladze:2006pa}.

In standard model, we have the perturbative U(1) gauge coupling at a
very high energy scale and the weak gravity conjecture predicts a
new intrinsic scale $\Lambda\sim \sqrt{\alpha/G_4}\sim 10^{17}$ GeV
which is much higher than the energy scale in colliders. If there
are new extremely weak gauge forces with coupling $g\sim 10^{-15}$,
a new scale for gauge theory at TeV scale appears. However it is
very difficult for us to detect such extremely weakly coupled gauge
theory. It seems that the weak gravity conjecture is irrelevant to
experiments.

On the other hand, the most exciting possibility raised by large
extra dimensions
\cite{Arkani-Hamed:1998rs,Antoniadis:1998ig,Arkani-Hamed:1998nn,Cremades:2002dh,Kokorelis:2002qi,Floratos:2006hs}
is that the fundamental Planck scale may be much lower than the
apparent four-dimensional Planck scale. This implies that we may
begin to experimentally access the dynamics of quantum gravity
sooner than previously anticipated.

In this short note, we propose that a new intrinsic UV cutoff for
U(1) gauge theory with large extra dimensions is proportional to the
fundamental Planck scale, not the four-dimensional Planck scale.
This new energy scale may be relevant for the physics at the LHC.

We compactify a d-dimensional theory to four dimensions, and the
four-dimensional Planck scale is determined by the fundamental
Planck scale $M_d$ in $d$ dimensions and the geometry of the extra
dimensions, \e M_4^2=M_d^{n+2}R^n, \label{mmd}\q where $n=d-4$ and
$R$ is the average size of the extra dimensions. If $M_d\sim 1$ Tev,
$R\sim 10^{13}$ cm for $n=1$ which is excluded, $R\sim 1$ mm for
$n=2$ which is roughly the distance where our present experimental
measurement of gravitational strength forces stops, and $R\sim
10^{-12}$ cm for $n=6$. The gravitational interaction is unchanged
over distances larger than the size of the extra dimensions $R$ with
the behavior of Newtonian potential $1/r$; however, the Newtonian
potential behaves as $1/r^{n+1}$ for $r\ll R$. In this scenario, the
standard model particles are always localized on a 3-brane embedded
in the higher-dimensional space. A review is given in
\cite{Arkani-Hamed:1998nn}.

The mass of a lightest black hole is larger than the fundamental
Planck scale; otherwise, a full quantum theory of gravity, such as
string theory, is needed. Since the fundamental Planck scale can be
much lower than the four-dimensional Planck scale, black holes not
too much heavier than the fundamental Planck scale may be produced
at lower energy scale \cite{Giddings:2001bu}. Following the idea in
\cite{Arkani-Hamed:2006dz}, we take the monopole mass as a probe of
the UV cutoff of a U(1) gauge theory in four dimensions. This U(1)
gauge theory arise in the Higgsing of an SU(2). The order of mass of
monopole is roughly \e M_{mon}\sim {\Lambda\over g^2},
\label{mass}\q if the theory has a cutoff $\Lambda$, and the size of
monopole is given by \e L_{mon}\sim {1\over \Lambda}.\label{size}\q
For electro-weak scale $10^2$ GeV, the size of monopole is roughly
$10^{-16}$ cm which is much smaller than the size of the extra
dimensions if $M_d\sim 1$ TeV. Here we want to stress that the scale
$1/R$ is the scale for the KK modes of the graviton, which means
that we should take the (n+4)-dimensional gravity into account above
the scale $1/R$. But this scale is not a scale for the gauge
theories on a 3-brane. We can easily check that the electro-weak
scale is higher than $1/R$. That is why we consider
(n+4)-dimensional gravity, not four-dimensional gravity. The gravity
for monopole propagates in $d$-dimensional space-time. The size of
black hole $R_{bh}$ in $d$ dimensions with mass $M$ takes the form,
\cite{Giddings:2001bu}, \e R_{bh}^{n+1}(M)\sim M_d^{-(n+2)}M. \q
Requiring that the monopole does not collapse to a $d$-dimensional
black hole, or equivalently $L_{mon}>R_{bh}(M_{mon})$, yields \e
\Lambda\leq g^{2\over n+2}M_d\sim \({g^2\over G_d} \)^{1\over n+2},
\label{wgc} \q where $G_d\sim M_d^{-(n+2)}$ is the Newton coupling
constant in $d$ dimensions. Otherwise black hole is contained and
this U(1) gauge theory is not an effective field theory. The new
intrinsic UV cutoff is proportional to the fundamental Planck scale,
not the four-dimensional Planck scale. Our result is different from
that in higher dimensional spacetime. In arbitrary dimensions, the
new UV cutoff predicted by weak gravity conjecture is proportional
to $g$ (see \cite{Banks:2006mm}), not $g^{2/(n+2)}$. Using eq.
(\ref{mmd}), we obtain \e \Lambda\leq {g^{2\over n+2}\over
(M_4R)^{n\over n+2}}M_4.\label{vgc} \q For $n=0$ which means there
is no extra dimension, eq. (\ref{wgc}) and (\ref{vgc}) are just the
same as (\ref{lgm}). In our note, we only pay attention to the case
with $n>0$. If the size of the extra dimension is much larger than
four-dimensional Planck length, i.e. $R\gg M_4^{-1}$, the constraint
on the UV cutoff in (\ref{wgc}) or (\ref{vgc}) is much more
stringent than that in (\ref{lgm}).

The above argument is consistent with the requirement that the
gravity should be the weakest force. We take the charged particle
with mass $m$ into account. The interaction between the charged
particles is described by a four-dimensional U(1) gauge theory. The
repulsive gauge force is roughly ${g^2\over r^2}$, where $r$ is the
distance between them. If $r$ is much smaller than the size of extra
dimensions, the gravitational force between them is ${G_d m^2\over
r^{n+2}}$. Weak gravity conjecture says ${g^2 \over r^2}\geq {G_d
m^2\over r^{n+2}}$, or \e m^2\leq {g^2 r^n \over G_d}. \label{wf}\q
This condition is easy to be satisfied if the charged particles are
separated far away from each other. On the other hand, the Compton
wavelength can be thought of as a fundamental limitation on
measuring the position of a particle, taking quantum mechanics and
special relativity into account. The reasonable distance between two
charged particles should be greater than the Compton wavelength of
the charged particle $m^{-1}$. Thus the most stringent constraint
occurs when $r\sim m^{-1}$. Now eq. (\ref{wf}) becomes \e m\leq
\({g^2\over G_d}\)^{1\over n+2}.\label{wgcc}\q Since the mass of the
charged particle is naively proportional to the UV cutoff $\Lambda$,
eq. (\ref{wgcc}) is just the same as (\ref{wgc}). So weak gravity
conjecture with large extra dimensions is nothing but the condition
for that gravity can be ignored.

According to eq. (\ref{wgc}), a new intrinsic UV cutoff for U(1)
gauge theory is roughly given by \e \Lambda\sim g^{2/(n+2)}M_d,\q or
a lower bound on the fundamental Planck scale takes the form \e
M_d\geq g^{-2/(n+2)}\Lambda.\q At electro-weak energy scale
$\Lambda_{ew}\sim 10^2$ GeV, U(1) coupling is roughly $g^2\sim
10^{-2}$. Since we didn't find any UV cutoff for U(1) gauge theory
under the electro-weak scale, the fundamental Planck scale should be
higher than $3\cdot 10^2$ GeV for $n=2$, or $2\cdot 10^2$ GeV for
n=6. If the fundamental Planck scale is $10$ TeV, weak gravity
conjecture predicts that a new intrinsic scale for U(1) gauge theory
is roughly $3$ TeV for $n=2$, or $6$ TeV for $n=6$. LHC with a
center of mass energy of $14$ TeV offers an opportunity to approach
the dynamics of quantum gravity and check weak gravity conjecture.

In this note, a much more stringent constraint on the effective
low-energy theories containing gravity and U(1) gauge fields is
obtained with large extra dimensions. Weak gravity conjecture
induces a new intrinsic UV cutoff for U(1) gauge theory in four
dimensions which is lower than the fundamental Planck scale.
Compactification with large extra dimensions offers a complete
natural understanding of the hierarchy in standard model if the
fundamental Planck scale is not much higher than $1\sim 10$ TeV. If
so, not only the black holes are possibly produced, but also a new
intrinsic UV cutoff lower than $10$ TeV for U(1) gauge theory in
standard model emerges. In standard model, U(1)$_{em}$ breaks down
above the electro-weak scale. There is an opportunity to test weak
gravity conjecture for U(1)$_Y$ at the LHC. We hope LHC bring us to
some surprising facts in the near future.

\vspace{.5cm}

\noindent {\bf Acknowledgments}

We would like to thank C. Leung, J.H. She, X. Wu and H. Yee for
useful discussions.

\end{document}